\documentclass[a4paper,10pt]{article}
\usepackage{graphicx}
\usepackage[labelfont={it,bf},font=it]{caption}
\usepackage{amsmath}
\usepackage{amsfonts}
\usepackage{subfigure}
\usepackage{pslatex}

\makeatletter
\renewcommand\section{\@startsection {section}{1}{\z@}%
    {-3.5ex \@plus -1ex \@minus -.2ex}%
    {2.3ex \@plus.2ex}%
    {\normalfont\bfseries\MakeUppercase}}
\makeatother

\newcommand{\be}{\begin{equation}}
\newcommand{\ee}{\end{equation}}
\newcommand{\bea}{\begin{eqnarray}}
\newcommand{\eea}{\end{eqnarray}}

\newcommand{\bfA}{\mathbf{A}}
\newcommand{\bfB}{\mathbf{B}}

\newcommand{\bfQ}{\mathbf{Q}}
\newcommand{\bfS}{\mathbf{S}}

\newcommand{\bfr}{\mathbf{r}}
\begin{document}
\setlength{\parindent}{0pt}
\pagestyle{empty}

\begin{center}
{\fontsize{16}{16pt} \bf
 $\bfS$ and $\bfQ$ Matrices Reloaded:\\
applications to open, inhomogeneous, and complex cavities}
\vspace{12pt}\vspace{12pt}\\
{\bf Guillaume Painchaud-April, Joey Dumont, Denis Gagnon, and Louis J. Dub\'e*}\\
{\it D\'epartement de Physique, de G\'enie Physique, et d'Optique, Facult\'e des Sciences et de G\'enie \\
Universit\'e Laval, Qu\'ebec, QC G1V 0A6, Canada\\
*Corresponding author: ljd@phy.ulaval.ca}
\end{center}
\vspace{-0.6cm} \vspace{6pt}
\section*{Abstract}

We present a versatile numerical algorithm for computing resonances of open dielectric cavities. The emphasis is on the generality of the system's configuration, i.e. the geometry of the (main) cavity (and possible inclusions) and the internal and external dielectric media (homogeneous and inhomogeneous). The method  is based on a scattering formalism to obtain the position and width of the (quasi)-eigenmodes. The core of the method lies in the scattering $\bfS$-matrix and its associated delay $\bfQ$-matrix which contain all the relevant information of the corresponding scattering experiment. For instance,
the electromagnetic near- and far-fields are readily extracted.   The flexibility of the propagation method is displayed for a selected system.

\section{Introduction}

In the research areas of microcavity physics \cite{vahala03},  lasing \cite{harayama11_LaserPhotonicsRev} and sensing 
\cite{vollmer12_Nanophoton}, increasing flexibility  in the geometries of the cavities and in the constituting/surrounding  media 
has been paramount for better performance regarding for instance, directional laser emission, lower lasing threshold and  higher 
sensing resolution. Confronted with this demand, we have set out to develop a versatile theoretical 
approach capable to meet the challenges of ever more complex experimental arrangements. Although finite element methods offered
themselves as a possible solution, we thought that they somewhat mask the underlying  physics. A scattering approach appeared to us
as the method of choice complementary to existing boundary element method (BEM) \cite{wiersig03} and boundary integral method 
(BIM) \cite{boriskina04a} to name a few. We have extended and upgraded a method \cite{rahachou03,rahachou04a} recently introduced
that has the potential to treat a wide variety of geometrical and refractive index deformations. This presentation is but a glimpse 
at the capabilities of the method. Details will appear elsewhere.

\section{Transfer, Scattering and Delay Matrices}

Our tale is that of a set of matrices: {\em transfer} matrices to propagate the solution from the interior of a system to the
outside world, a {\em scattering} $\bfS$-matrix to collect the information and a derived {\em delay} $\bfQ$-matrix to provide the eigen-modes of the system and their quality factors (or delays).  
There seems to have been renewed interest of late on the properties of the $\bfQ$ matrix in different contexts, and we refer the reader to some recent publications for further details \cite{chong11_prl,rotter11_prl,schomerus10_prl,shimamura11_jpb}.

\begin{figure}[h]
\centering
\subfigure[][ \ Concentric domains]{\includegraphics[width = 6cm]{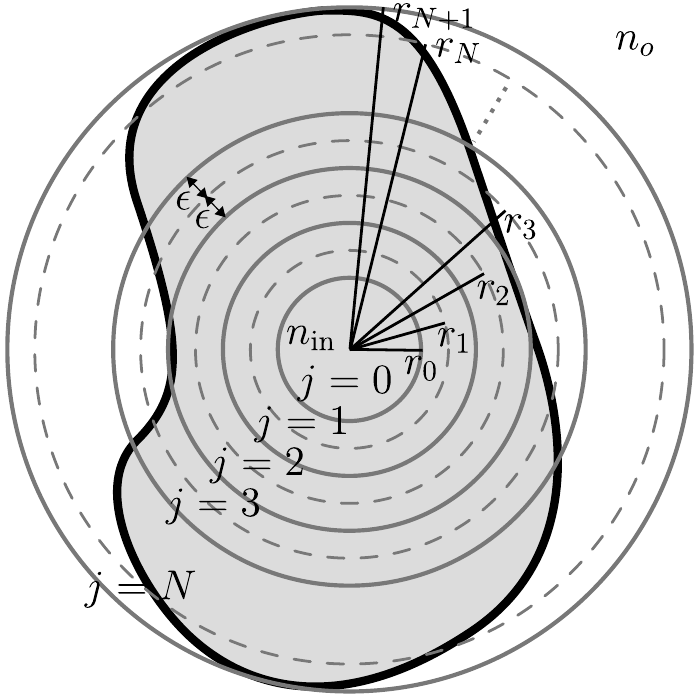} \label{fig1a}} \hfill
\subfigure[][ \ Expansion coefficients]{\includegraphics[width = 6cm]{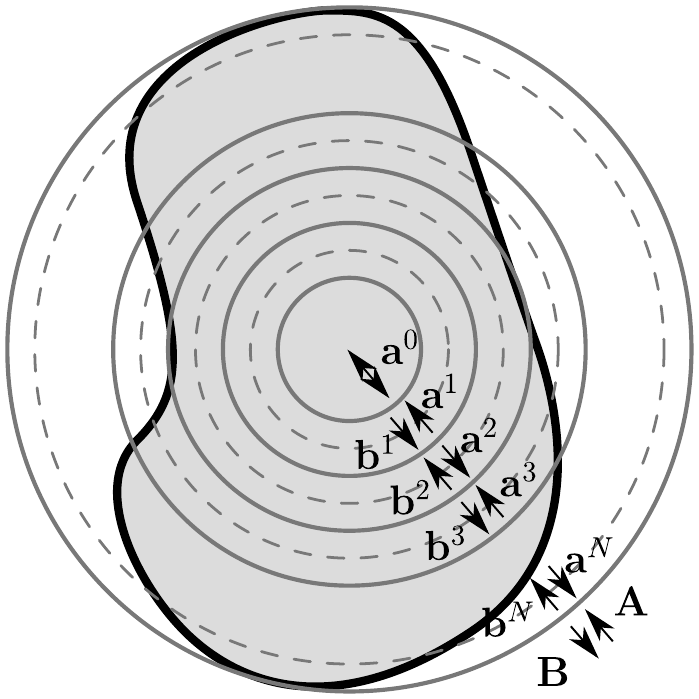}\label{fig1b}}
\caption{\subref{fig1a} Schematic representation of the separation of a generic cavity in annular domains. The refractive indices,
$n_\mathrm{in}$ and $n_o$, are constant within the innermost disk and outside of the outermost domain respectively.
\subref{fig1b} The expansion coefficients $\mathbf{a}^j$ and  $\mathbf{b}^j$ at the successive concentric domains are used to propagate the solution to the exterior region where the $\bfS$ matrix ($\bfB= \bfS \bfA$) is obtained.
 \label{fig1_domains+coeff}}
\end{figure}

Here is a brief summary of the important elements of the method. For the sake of simplicity, let us consider the solution 
$\psi(\bfr)$ of an electromagnetic wave in a two-dimensional (2D) dielectric medium. We will concentrate on TM polarisation. The equation to solve is the Helmholtz equation
\be
       \left[  \nabla^2 + n^2(\mathbf{r}) k^2 \right] \psi(\mathbf{r}) = 0
        \label{eqn:HelmholtzEqn}
\ee
which, in polar coordinates, takes the form
\begin{equation}
    \left[r^2\frac{\partial^2}{\partial r^2}+r\frac{\partial }{\partial
    r}+ \frac{\partial^2}{\partial \phi^2} +
    n^2(r,\phi)k^2r^2\right]\psi(r,\phi) = 0 .\label{eq2_3_1}
\end{equation}
We then proceed to divide space into $N+1$ interfaces ($j= 0, 1, \ldots, N$) located at distances $R_j= r_0 +2j\epsilon$
(Fig. \ref{fig1a}). To follow the procedure, let us focus our attention on a thin annular region of thickness $2\epsilon$ and central radius $r_j$, $j=1,2...N$, covering part of the generic cavity. Over this region, we assume the refractive index\footnote{
The refractive index is real in the present description. However, the method can easily accommodate a complex refractive index
to account for absorption and loss processes. }
 to be dependent upon $\phi$ only and $n^2(r,\phi)r^2$ is then evaluated at $r_j$ for all $\phi$. Turning to (\ref{eq2_3_1}), we obtain an approximate local expansion of the differential equation over $|r-r_j| \leq \epsilon$,
\begin{equation}
\left[r^2\frac{\partial^2}{\partial r^2}+r\frac{\partial }{\partial     r}+ \frac{\partial^2}{\partial \phi^2} +
    n^2(r_j,\phi)k^2r_j^2\right]\!\psi^j(r,\phi) =  0 .\label{eq2_3_2}
\end{equation}
This differential equation is separable for the local wavefunction $\psi^j(r,\phi)$. Under the Ansatz $\psi^j(r,\phi) = \mathcal{R}^j(r)\Phi^j(\phi)$, we obtain two exact differential equations
\begin{eqnarray}
    \left[\rho_j^2\frac{d^2}{d\rho_j^2}+\rho_j\frac{d}{d
    \rho_j} - \xi^j\right]\mathcal{R}^j(\rho_j) & = &
    0 \label{eq2_3_3}\\
    \left[\frac{d^2}{d\phi^2}+\left(n^2(r_j,\phi)k^2r_j^2 + \xi^j\right) \right]\Phi^j(\phi) & = &
    0 \label{eq2_3_4}
\end{eqnarray}
where $\rho_j=r/r_j$ and $\xi^j$ is a separation constant. Imposing periodic boundary conditions on Eq. (\ref{eq2_3_4}) and expanding $\Phi^j(\phi)$ in a Fourier series, we end up with an eigenvalue problem of an hermitian symmetric matrix. The retrieved eigenvalues specify the set of constants $\{\xi_\mu^j\}$, and the associated eigenvectors define a basis of normalized eigenfunctions $\{\Phi^j_\mu(\phi)\}$. We may then solve the Cauchy-Euler differential equation (\ref{eq2_3_3}) {\em exactly},
\begin{equation}
    \mathcal{R}_\mu^j(\rho_j) =
    a_\mu^j\rho_j^{+\sqrt{\xi_\mu^j}}+b_\mu^j\rho_j^{-\sqrt{\xi_\mu^j}}\label{eq2_3_4+1}
\end{equation}
where $a_\mu^j$ and $b_\mu^j$ are constants to be obtained from the boundary conditions. The sign of $\xi_\mu^j$ renders $\mathcal{R}_\mu^j(\rho_j)$  evanescent (- sign) or propagating (+ sign) (Fig. (\ref{fig1b}). 
           
This procedure is propagated to an adjacent thin annular domain of thickness $2\epsilon$ centered on $r_{j+1} = r_j+2\epsilon$. Boundary conditions for TM polarisation assume continuity of $\psi^j$ and $\psi^{j+1}$ and their normal (radial) derivative at their common boundary $\rho_{j+} = 1+\epsilon/r_j$ and $\rho_{j+1-} = 1-\epsilon/r_{j+1}$. 

Finally, having reached the outermost region ($r \ge R_{max}= r_N + \epsilon$), the solution is matched to a 
superposition of {\em incoming} $H^{(-)}$ and {\em outgoing} $H^{(+)}$ Hankel functions
\begin{equation}
    \psi(r,\phi) = \sum_m\left[A_mH^{(-)}_m(n_okr)+B_mH^{(+)}_m(n_okr)\right]\mathrm{e}^{im\phi}  ,\label{eq2_3_10}
\end{equation}
 from which the scattering $\bfS$-matrix is defined
\be
              \bfB = \bfS \, \bfA .
\ee
One can then show, and this is a crucial part of the approach, that the characteristic expansion vectors $\bfA$ are nothing less than
the eigenvectors $\mathbf{A}^{p}$ of the delay $\bfQ$-matrix \cite{shimamura11_jpb,smith60} given by     
\be
              \bfQ =  -i \mathbf{S}^\dag\frac{\partial \mathbf{S}}{\partial k} =
                          i  \frac{\partial \mathbf{S}^\dag}{\partial k} \mathbf{S}
                \label{eq2_Qmatrix}
\ee
such that
\be
            \bfQ \mathbf{A}^{p} = c\tau_p \mathbf{A}^{p}
 \ee
where $c \tau_p$ (in length units) are the eigen-delays of a characteristic mode $p$ proportional to its quality factor or inversely proportional to its (quasi)-resonance width.   
              
Clearly, this short description does not do honour to the daunting task of implementing the propagation steps, transferring information between adjacent annular domains, nor does it do justice to the theoretical subtleties necessary  to interpret the results obtained. 
However, it should suffice to give the jest of the method. An example will give a glimpse of its capabilities..
                  
\section{Results: A Selected Example}
\begin{figure}[h!]
\begin{minipage}[b]{0.55\linewidth}
\centering
\subfigure[]{\includegraphics[width=0.50\textwidth]{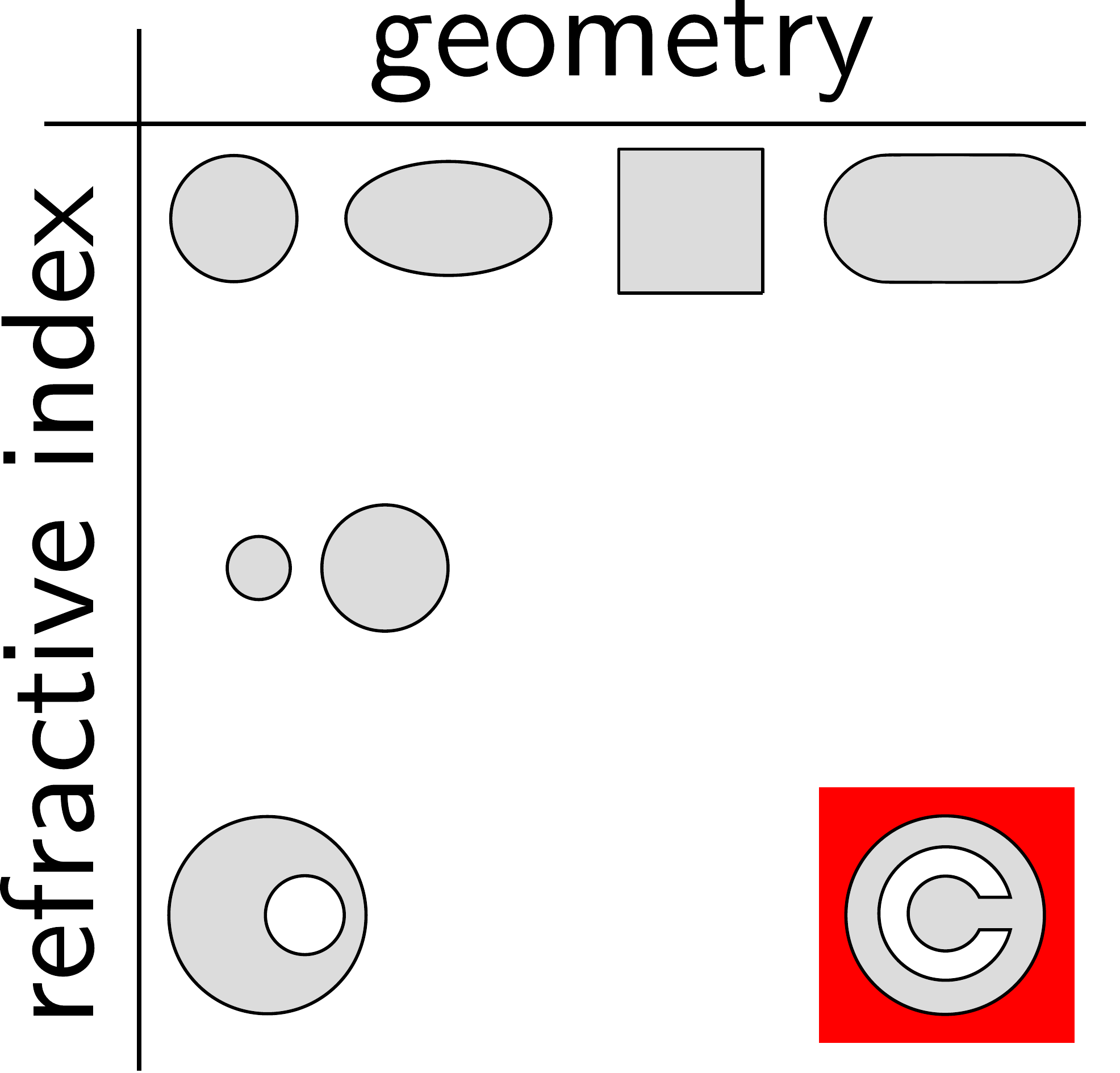}\label{fig2a_geometries}} \\ \vfill
\subfigure[]{\includegraphics[width=0.80\textwidth]{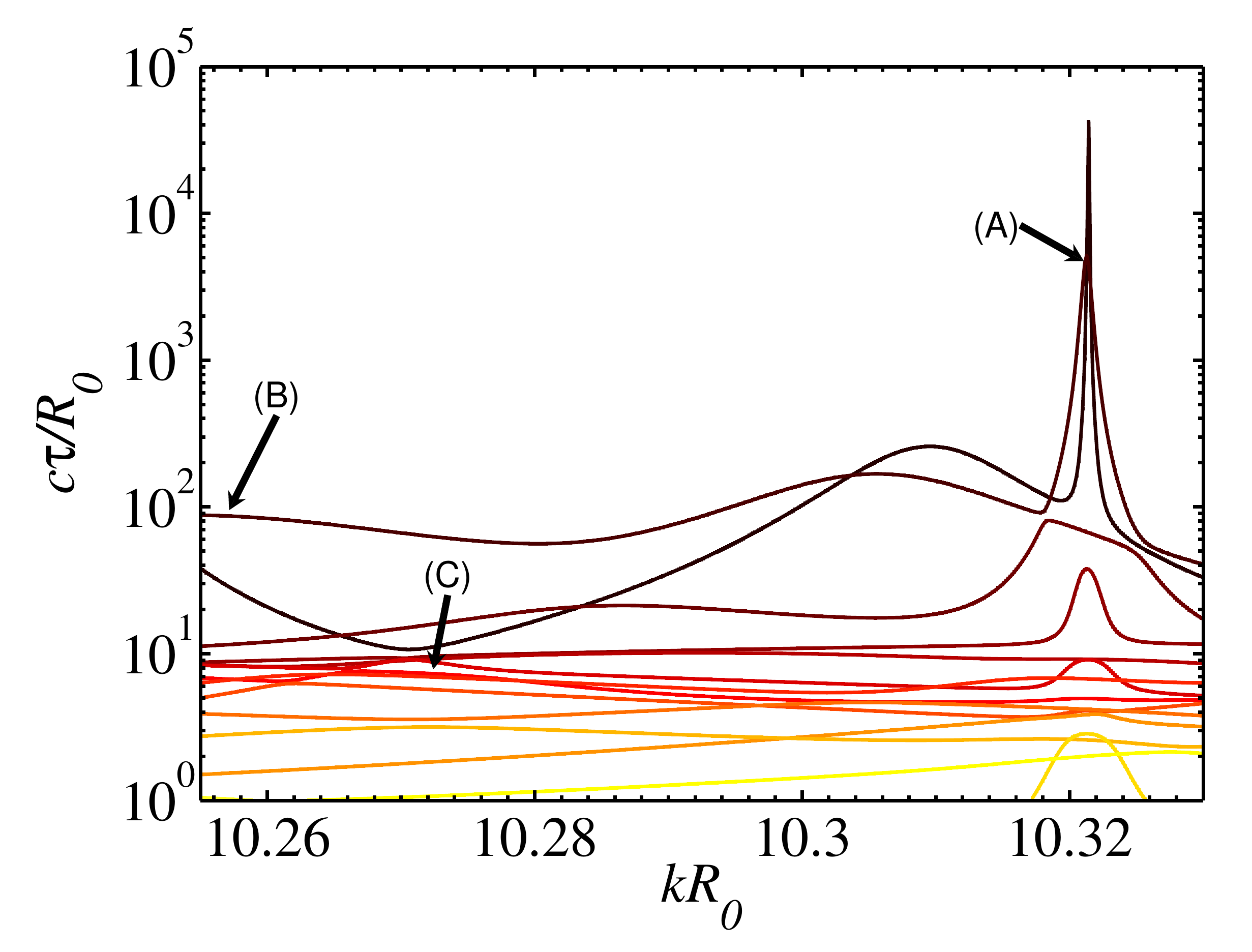}\label{fig2b_spectrum}} 
\end{minipage}
\begin{minipage}[b]{0.55\linewidth}
\centering
\setcounter{subfigure}{0}
\subfigure{\includegraphics[width=0.4\textwidth]{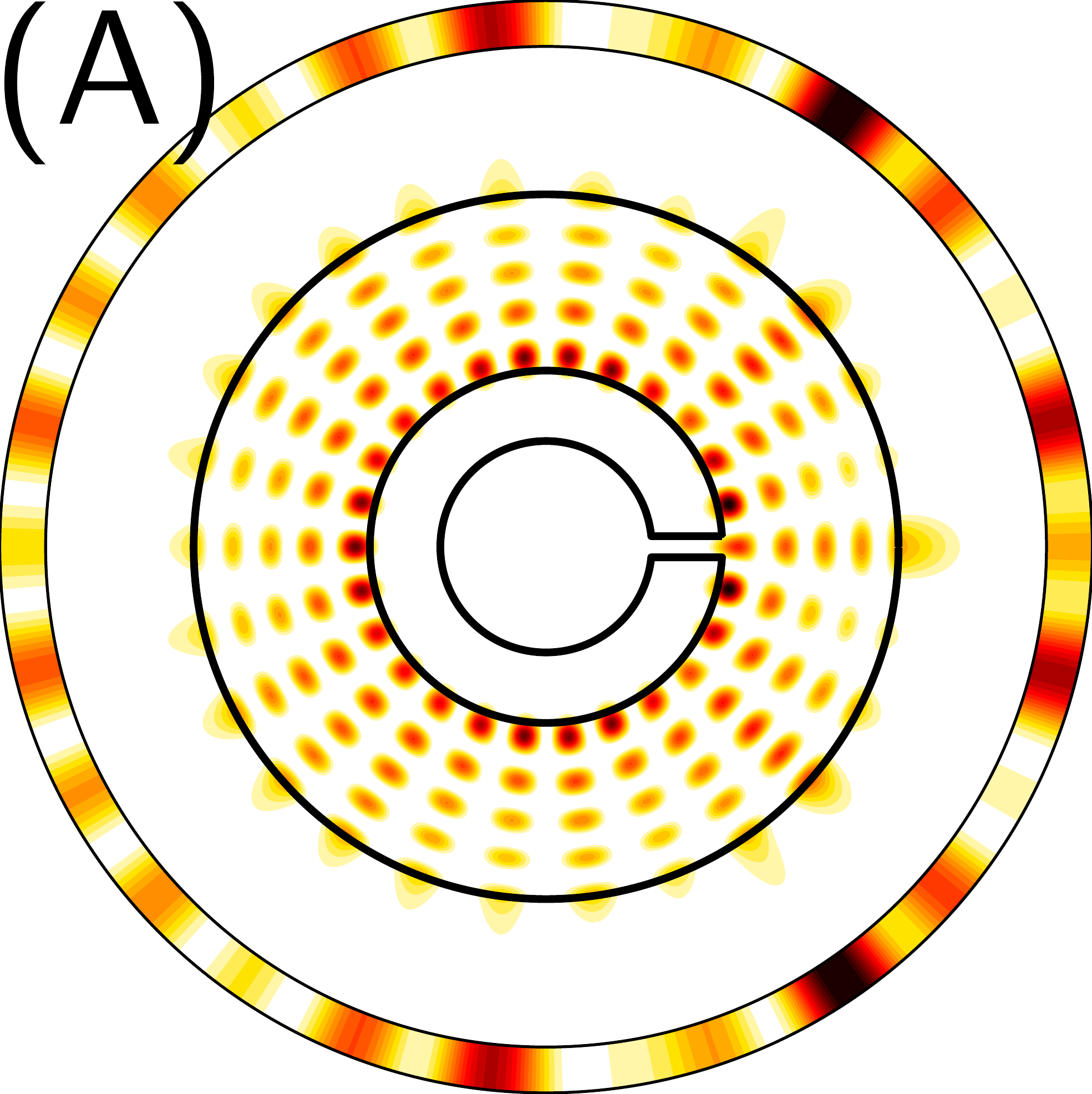}} \\
\subfigure{\includegraphics[width=0.4\textwidth]{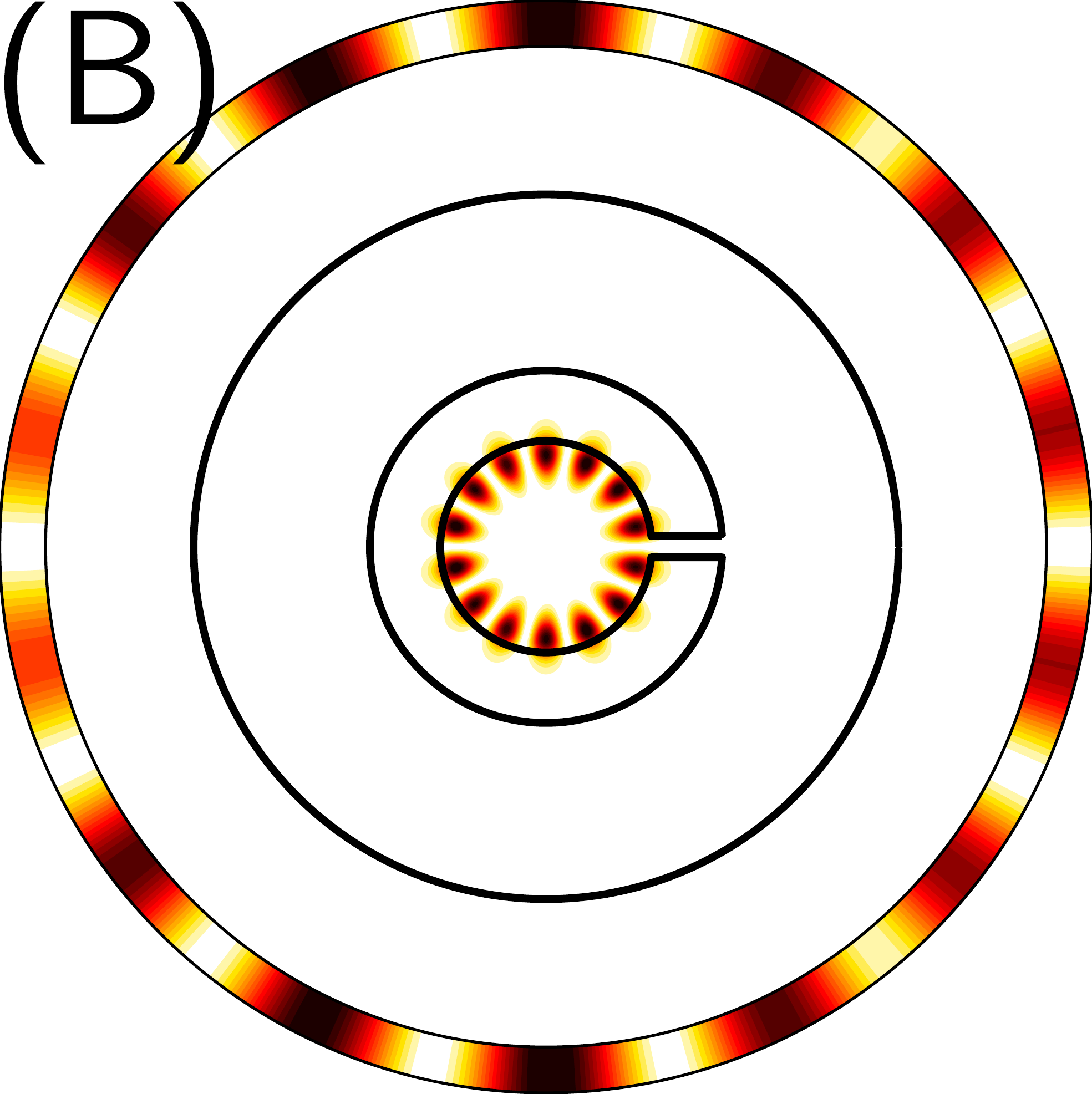}} \\
\subfigure[]{\includegraphics[width=0.4\textwidth]{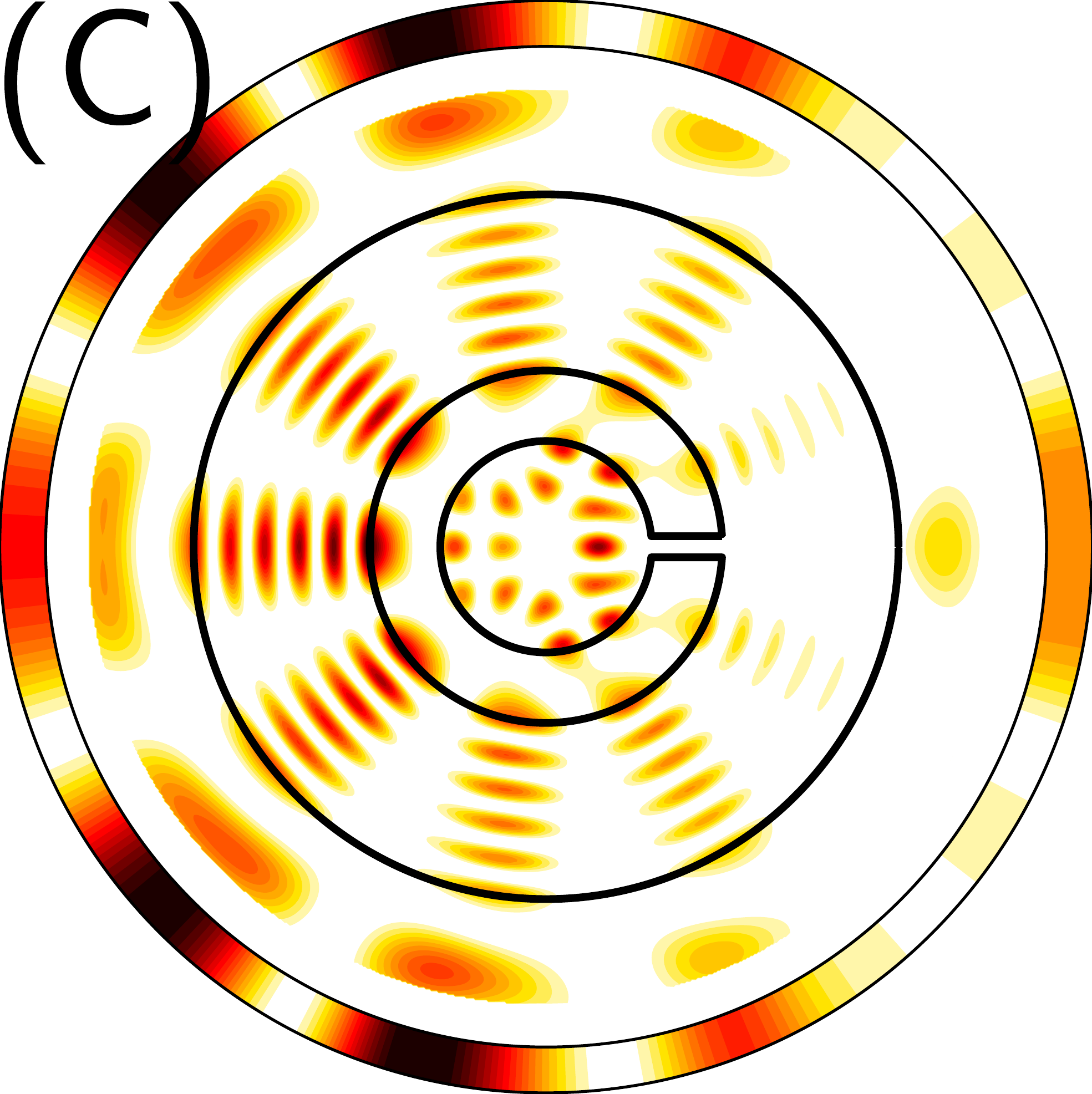}\label{fig2c_distribution}}
\end{minipage}
\caption{\subref{fig2a_geometries} Typical arrangements of geometrical distortions and/or inhomogeneous dielectric media, 
the combined complexity of the systems increase along the diagonal.
\subref{fig2b_spectrum} Representative eigen-delay spectrum for the $C$ inclusion.
\subref{fig2c_distribution} Intensity distribution of the near-field (NF) and the far-field (FF) in false colour (the darker the colour,
the higher the intensity).  The outer-circle serves as a support to display the FF.} \label{fig2_inclusion}
\end{figure}

Figure (\ref{fig2a_geometries}) present some of the geometries that we have used for the calibration of the method: on the top row,
we see a number of homogeneous cavities of increasing distortions with respect to a perfect disk, while the discontinuities in the dielectric media  increase from top to bottom. We present in Figs (\ref{fig2b_spectrum} -\ref{fig2c_distribution}) 
results of a calculation of a more complex
arrangement with both a geometrical distortion and  a dielectric inclusion (in form of a $C$, for complex no less). The sweep over a
wavenumber window (Fig. \ref{fig2b_spectrum}) reveals a number of resonances of different quality factors: the ordinate axis displays the eigenvalues of the $\bfQ$ matrix and are directly proportional to the width of the corresponding resonances. 
In Fig. (\ref{fig2c_distribution}),
we show the near- and far-field distribution for 3 selected resonances appearing in the spectrum.  They have been chosen for their
respective differences. Resonance (A) has the near-field (NF) structure of a whispering gallery type, 
almost uniform, whereas its far-field 
(FF) distribution shows pronounced peaks at certain privileged directions. Resonance (B) has a NF amplitude completely located within the inclusion
and a FF with a dominant distribution in a direction perpendicular to the opening of the inclusion. Resonance (C), a low quality mode, has for its part strongly non-uniform NF {\em and} FF distributions. 

Typically, our method handles well complicated combinations of various geometries and medium discontinuities. However, it is fair to say that its careful implementation has only been thoroughly tested for two-dimensional systems. Among other things, the three-dimensional cases are waiting in the wings ...  
%

\section{Conclusion}
We have presented a scattering/propagation method applicable to cavities of arbitrary shape and arbitrary inhomogeneities of the medium (continuous and/or discontinuous). The calculational approach has now reached maturity for 2D systems, and is currently being 
extended to photonic molecules (combination of a few cavities), and more generally, to photonic complexes (periodic or aperiodic arrangements of cavities). Details of the implementation, and further results will be published elsewhere. 

The authors acknowledge financial support from the Natural Sciences and Engineering Research Council of Canada (NSERC) and computational resources from Calcul Qu\'ebec. J.D. is grateful to the Canada Excellence Research Chair in Enabling Photonic Innovations for Information and Communication for a research fellowship and personally thanks Y. Messaddeq and J.-F. Viens.


\end{document}